\documentclass[review]{elsarticle}
\usepackage{textcomp}
\usepackage{graphicx}
\usepackage{subcaption}
\usepackage{epstopdf}
\usepackage{adjustbox}
\usepackage{booktabs}
\usepackage{placeins}
\usepackage{algorithm,algorithmic}
\usepackage{multirow}
 \usepackage[normalem]{ulem}
 \useunder{\uline}{\ul}{}
\journal{Journal of \LaTeX\ Templates}

\begin{document}

\begin{frontmatter}

\title{Stacked Convolutional Neural Network for Diagnosis of COVID-19 Disease from X-ray Images}


\author{\large Mahesh Gour $^{1^*}$, Sweta Jain $^{2}$}
\address{$^{*}$corresponding author: maheshgour0704@gmail.com\\
$^{1,2}$ Maulana Azad National Institute of Technology, Bhopal, MP, India 462003}

\begin{abstract}
Automatic and rapid screening of COVID-19 from the chest X-ray images has become an urgent need in this pandemic situation of SARS-CoV-2 worldwide in 2020. However, accurate and reliable screening of patients is a massive challenge due to the discrepancy between COVID-19 and other viral pneumonia in X-ray images. In this paper, we design a new stacked convolutional neural network model for the automatic diagnosis of COVID-19 disease from the chest X-ray images. We obtain different sub-models from the VGG19 and developed a 30-layered CNN model (named as CovNet30) during the training, and obtained sub-models are stacked together using logistic regression. The proposed CNN model combines the discriminating power of the different CNN\textquotesingle s sub-models and classifies chest X-ray images into COVID-19, Normal, and Pneumonia classes. In addition, we generate X-ray images dataset referred to as COVID19CXr, which includes 2764 chest x-ray images of 1768 patients from the three publicly available data repositories. The proposed stacked CNN achieves an accuracy of 92.74\%, the sensitivity of 93.33\%, PPV of 92.13\%, and F1-score of 0.93 for the classification of X-ray images. Our proposed approach shows its superiority over the existing methods for the diagnosis of the COVID-19 from the X-ray images. 
\end{abstract}
\begin{keyword}
    COVID-19 \sep automatic screening \sep stacked generalization \sep ensemble technique \sep deep learning \sep logistic regression \sep chest X-ray images. 
\end{keyword}

\end{frontmatter}

%\linenumbers

\section{INTRODUCTION}
The Novel coronavirus disease 2019 (COVID-19) pandemic has put the livelihoods and health of the massive population in a critical position. It has led to the disturbance throughout the public life of the world population. The severe acute respiratory syndrome coronavirus 2 (SARSCoV-2) belongs to the family of Coronavirus, which get transmitted in the people based on the infection in the form of direct contact or fomites. The primary symptoms of coronavirus infection are fever, cough and fatigue. In several cases coronavirus cause severe respiratory problems like Pneumonia, lung disorders and kidney malfunction. The virus has serious consequences as its serial interval is 5 to 7.5 days, and the rate of reproduction is 2 to 3 \cite{nishiura2020serial} people.  The coronavirus infection can incite SARS (Severe Acute Respiratory Syndrome), which might unfold serious health impacts.

It is estimated that many people are healthy carriers of a virus, and they are the reason for about 5\% to 10\% of acute respiratory infections  \cite{chen2020emerging}. A critical step to fight against the COVID-19 is to identify the infected people so that they get immediate treatment and isolate them to control the multiplying of the spread of the infection. 

The COVID-19 panic has increased due to the unavailability of fast and accurate diagnosis systems to test the infected people. According to the World Health Organization (WHO), the diagnosis of COVID-19 cases must be  confirmed by molecular assay, such as the reverse transcription polymerase chain reaction (RT-PCR) pathological test using throat swab samples \cite{world2020novel}. While RT-PCR has become a standard tool for confirmation of COVID-19, but it is a very time consuming, laborious, and manual process, and there is a limitation of availability of diagnostic kits and sample collection. The availability of COVID-19 testing kits is limited as compared to the increasing amount of infected people; hence there is a need to rely on different diagnosis methodologies.

The coronavirus  targets the epithelial cells that affect patients respiratory tract, which can be analyzed by the radiological images of a patient’s lungs. Some early studies also show that patients present anomalies in chest x-ray images, which are the typical characteristics of COVID-19 infected patients \cite{ng2020imaging, huang2020clinical}. Hence, the development of the computer-aided diagnosis system for the automatic analysis of radiological images can be very helpful in identifying infected patients at a faster rate. Some the advantage of the using X-ray images for COVID-19 screening as follows: 
\begin{itemize}
    \item Enable fast screening and rapid triaging of patients suspected of COVID-19.
    \item 	Making use of readily available and accessible radiological images. 
    \item Portable and easy to setup, these systems can be setup in the isolation room, which significantly minimizes the risk of transmission. 
\end{itemize}

Recent advancements in deep learning specifically in Convolution Neural Network (CNN) motivated us to develop a highly reliable computer-aided diagnosis (CAD) systems 
for the rapid detection of the COVID-19 using chest X-ray images. It will enable and enhance the automation in the screening phase, which is the crucial part of the COVID-19 pandemic. 
In this study, we design a new stacked convolutional neural network for automatic diagnosis of COVID-19 disease from the chest X-ray images. To train the proposed model, we generated chest X-ray images dataset, with the combination and modification of three publicly available datasets \cite{Chung2020covid, cohen2020covid, kermany2018labeled}, which we will refer to as COVID19CXr.

\par The organization of this paper as follows: Section 2 presents the related work.  Section 3 describes the proposed stacked  CNN model for the classification of COVID-19, Normal, and Pneumonia X-ray images. Section 4 describes the COVID19CXr dataset generation process and details the experimental results, performance comparison. Finally, the conclusion is drawn in Section 5. 

\section{RELATED WORK}
Over the past 40 years, many computer-aided systems have been developed for the diagnosis of lung disease \cite{doi2007computer} and achieved promising results for automatic detecting lung abnormality from the radiological images \cite{castellano2004texture, van2002automatic,jaeger2013automatic}. 

Recently, automatic CAD of COVID-19 using radiological images has drawn a lot of attention of researchers and as a result several approaches have been introduced in literature. 
They have published a series of research articles \cite{shi2020review,dong2020role} demonstrating the CAD systems for the detection of COVID-19 using radiological images. Butt at el. \cite{butt2020deep} have studied various CNN models technically and proposed a model with the combination of 2D and 3D CNN models for the classification of the CT images into COVID-19, Influenza viral pneumonia, or no-infection. Their approach achieved a sensitivity of 98.2 \% and specificity of 92.2 \%. Ardakani et al. \cite{ardakani2020application} have presented the application of deep learning in COVID-19 detection using CT images. Authors tested ten pre-trained CNN models namely AlexNet, MobileNet-V2 VGG-16, VGG-19, ResNet-18, ResNet-50, ResNet-101, SqueezeNet, GoogleNet, and Xception. Their experiment results showed that ResNet101 performed best with area under the curve (AUC) of 0.99 over 1020 CT images of 194 patients. 

Similarly, Narin et al. \cite{narin2020automatic} have applied ResNet50, InceptionV3 and Inception-ResNetV2 using transfer learning for classification of the X-ray images into normal and COVID-19 class. This method achieved good performance with an accuracy of 98 \% with ResNet50. However, the number of X-ray images are only 100 and the number of images was very less. Wang et al. \cite{wang2020covid} have proposed an open-source COVID-Net model based on the projection-expansion-projection design pattern for COVID-19 cases detection from the X-ray images. In this study, the authors reported an accuracy of 92.6 \%. 
Oh et al. \cite{oh2020deep} have proposed patch-based CNN approach, to train ResNet18 model using image patches that have been extracted from the chest 
x-ray images. For decision making, they used the majority voting strategy, which resulted in an accuracy of 88.9 \%. 

An objected detection based DarkCovidNet model has been proposed by Ozturk et al. \cite{ozturk2020automated} for automatic detection of COVID-19 cases from the X-ray images. They have reported accuracy of  98.08 \% for binary classification of X-ray images into COVID-19 and no-findings, and also this approach has achieves an accuracy of 87.02 \% for multi-class classification of X-ray images into COVID-19, no-findings and pneumonia. Pereira et al.  \cite{pereira2020covid} have proposed a hierarchical classification approach, in which they extracted deep features by InceptionV3 and tested texture descriptors. They investigated early and late fusion techniques for combining the strength of descriptor and classifiers. Their hierarchical classification approach achieved F1-Score of 0.89 for the
COVID-19 identification in the X-ray images.         

Attention-based deep 3D multiple instance learning approach has been proposed by the Han et al. \cite{wang2020prior} for automatic screening of COVID-19 from the CT images. Their algorithm achieved an accuracy of 97.9\%. Wang et al. \cite{hu2020weakly} developed a weakly supervised deep learning framework, in which lung region was segmented by UNet from the CT images and  3D deep neural network is applied on the segmented region for predicting probability of COVID-19 infections. Authors reported an accuracy of 90.10\%.     

Ucar et al. \cite{ucar2020covidiagnosis} proposed a SqueezeNet CNN model with Bayesian optimization for diagnosis of the COVID-19 from X-ray images. They reported an accuracy of 98.26 \%. Afshar et al. \cite{afshar2020covid} developed a capsule network-based framework for the classification of the X-ray images into Normal, bacterial, Non-COVID, and COVID-19 cases. The authors reported an accuracy of 95.7 \% and a sensitivity of 90 \%. Sethy et al. \cite{sethy2020detection} extracted deep features of X-ray images from the pre-trained CNN, and support vector machine (SVM) has been applied on the extracted feature to classify x-ray images. The authors achieved an accuracy of 95.38\% using ResNet50 with the SVM classifier.

\section{METHODOLOGY}

The Convolution Neural Network is the driving concept of deep learning algorithms in computer vision, which led to outstanding performance in most of the pattern recognition tasks such as image classification \cite{krizhevsky2012imagenet,he2016deep,simonyan2014very}, object  localization, segmentation and detection \cite{girshick2014rich,girshick2015fast,gour2019deeprnnetseg}. It has also shown its superiority in the medical image analysis for image classification, and segmentation problem \cite{ronneberger2015u,gour2020residual, anwar2018medical, shen2017deep}, especially in lung-related diseases such as lung nodule detection \cite{huang2017lung}, pneumonia detection \cite{rajpurkar2017chexnet}, and pulmonary tuberculosis \cite{liu2017tx}. CNN automatically learns a low to the high level of useful feature representations and integrates feature extraction and classification stages in a single pipeline, which is trainable in an end-to-end manner without requiring any manual design and expert human intervention.

 \begin{figure}[!b]
        \centering
        \includegraphics[width=1.3\linewidth]{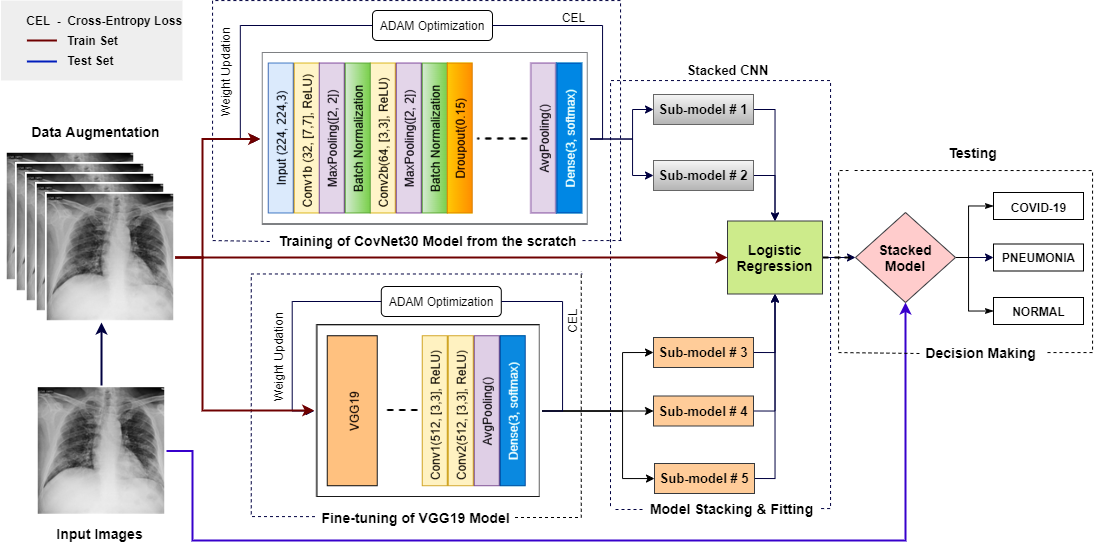}
        \caption{Block diagram of proposed Stacked CNN model}
\end{figure}%
In this work, we have developed a deep learning-based   stacked convolutional neural network for the rapid screening of COVID-19 patients using X-ray images. The proposed COVID-19 detection method includes three modules as shown in the Figure 1. In the first module, a new 30-layered CNN model is built and trained on the chest X-ray images from scratch, which will be referred to as CovNet30 (\textbf{Cov}id-19 \textbf{Net}work of \textbf{30}-layers). In the second module, a pre-trained VGG19 model \cite{krizhevsky2012imagenet} is fine-tuned on X-ray images for the diagnosis of COVID-19 disease. Finally, in the last module, five CNN\textquotesingle s sub-models are obtained, during the training of CovNet30 and VGG19 models. The  output of CNN\textquotesingle s  sub-models are stacked together by applying logistic regression \cite{breiman1996stacked} for building a new final CNN model for diagnosis of COVID-19 disease from X-ray images.
A detailed description of the proposed approach is given in the following section:
\subsection{CovNet30 Architecture and Training}
The CovNet30 is a task-specific, 30-layered convolutional neural network for X-ray images classification. It learns the non-linear, discriminative features directly from the chest X-ray images at multiple levels of abstraction. A detailed layer configuration of the CovNet30 network is shown in Table 1. CovNet30 is sequential network, in which Convolutional layer with ReLU activation, pooling layer, Batch Normalization layer and dropout layer are added repetitively. Every layer of CovNet30 produce a volume of activation to the next layer through  a differentiable function. Description of layers are given as follows: 
\begin{itemize}
    \item Convolution Layer extracts features from the input volume by performing convolution operation. In the convolution operation, the dot product is computed between the kernels and connected local regions of the input volume of activation.
    \item ReLU activation function is an elementwise activation function. If \textit{x} is input of ReLU then it produce\textit{ max(0, x)} (non-negative value) in the output.
    \item Pooling layer down sampling feature maps  and reduces the computation in the network.  
    \item  Batch Normalization layer speed up the learning process and improves network stability by minimizing the internal covariate shift. 
    \item Dropout is a regularization method, which prevents the network from the overfitting by dropping out units in the network. 
     
\end{itemize}

\begin{table}[!h]
\caption{Detailed layer configuration of CovNet30 network}
\begin{adjustbox}{max width=\textwidth}
\begin{tabular}{llccc} \hline
\multicolumn{1}{c}{\textbf{Layer   Name}} & \multicolumn{1}{c}{\textbf{Type}} & \textbf{Kernal Size} & \textbf{Output} & \textbf{Parameters} \\ \hline
conv2D\_1       & Convolutional + ReLU      & 7 {$\times$} 7 & 218 {$\times$} 218 {$\times$} 32 & 4736    \\
max\_pooling\_1  & Max Pooling               & 2 {$\times$} 2 & 109 {$\times$} 109 {$\times$} 32 & 0       \\
batchNo\_1      & Batch Normalization       & -     & 109 {$\times$} 109 {$\times$} 32 & 128     \\
conv2D\_2       & Convolutional + ReLU      & 5 {$\times$} 5 & 105 {$\times$} 105 {$\times$} 64 & 51264   \\
max\_pooling\_2  & Max Pooling               & 2 {$\times$} 2 & 52 {$\times$} 52 {$\times$} 64   & 0       \\
batchNo\_2      & Batch Normalization       & -     & 52 {$\times$} 52 {$\times$} 64   & 256     \\
dropout\_1      & Dropout                  & -     & 52 {$\times$} 52 {$\times$} 64   & 0       \\
conv2D\_3       & Convolutional + ReLU      & 3 {$\times$} 3 & 50 {$\times$} 50 {$\times$} 128  & 73856   \\
max\_pooling\_3  & Max Pooling               & 2 {$\times$} 2 & 25 {$\times$} 25 {$\times$} 128  & 0       \\
batchNo\_3      & Batch Normalization       & -     & 25 {$\times$} 25 {$\times$} 128  & 512     \\
dropout\_2      & Dropout                  & -     & 25 {$\times$} 25 {$\times$} 128  & 0       \\
conv2D\_4       & Convolutional + ReLU      & 3 {$\times$} 3 & 23 {$\times$} 23 {$\times$} 128  & 147584  \\
max\_pooling\_4  & Max Pooling               & 2 {$\times$} 2 & 11 {$\times$} 11 {$\times$} 128  & 0       \\
batchNo\_4      & Batch Normalization       & -     & 11 {$\times$} 11 {$\times$} 128  & 512     \\
dropout\_3      & Dropout                  & -     & 11 {$\times$} 11 {$\times$} 128  & 0       \\
conv2D\_5       & Convolutional + ReLU      & 3 {$\times$} 3 & 9 {$\times$} 9 {$\times$} 256    & 295168  \\
batchNo\_5      & Batch Normalization       & -     & 9 {$\times$} 9 {$\times$} 256    & 1024    \\
dropout\_4      & Dropout                  & -     & 9 {$\times$} 9 {$\times$} 256    & 0       \\
conv2D\_6       & Convolutional + ReLU      & 3 {$\times$} 3 & 7 {$\times$} 7 {$\times$} 256    & 590080  \\
batchNo\_6      & Batch Normalization       & -     & 7 {$\times$} 7 {$\times$} 256    & 1024    \\
dropout\_5      & Dropout                  & -     & 7 {$\times$} 7 {$\times$} 256    & 0       \\
conv2D\_7       & Convolutional + ReLU      & 3 {$\times$} 3 & 5 {$\times$} 5 {$\times$} 512    & 1180160 \\
batchNo\_7      & Batch Normalization       & -     & 5 {$\times$} 5 {$\times$} 512    & 2048    \\
dropout\_6      & Dropout                  & -     & 5 {$\times$} 5 {$\times$} 512    & 0       \\
conv2D\_8       & Convolutional + ReLU      & 3 {$\times$} 3 & 3 {$\times$} 3 {$\times$} 512    & 2359808 \\
batchNo\_8      & Batch Normalization       & -     & 3 {$\times$} 3 {$\times$} 512    & 2048    \\
dropout\_7      & Dropout                  & -     & 3 {$\times$} 3 {$\times$} 512    & 0       \\
globAvgPooling & Global AvgPooling         & -     & 512            & 0       \\
FC\_1           & Fully Connected + ReLU    & -     & 1000           & 513000  \\
FC\_2           & Fully Connected + Softmax & -     & 3              & 3003   \\ \hline
\end{tabular}
\end{adjustbox}
\end{table}

CovNet30 has been trained on the X-ray images in a supervised manner. Cross-entropy loss function is used to calculate the training error and which is minimize using the ADAM optimizer \cite{kingma2014adam}. Cross-entropy loss function is mathematical represented in equation (1).

\begin{equation} 
 J(T, P)=-\sum_{i=1}^{C} t_{i} \log \left(p_{i}\right)
 \label{eqn:loss}
 \end{equation}
 Where $t_{i}$ and $p_{i}$ are the target value and predicted probability for each class i in C. 

In the experiment, the values of hyper-parameter are set as follows:  learning rate to 0.001, the batch size to 16, and dropout probability to 0.15. 
we experimentally find that these are the best suitable values of hyper-parameters for network training. 

\subsection{Fine-Tuning of VGG19}
The VGG19 is a pre-trained network that is trained on the ImageNet dataset, which achieved state-of-the-art performance on ILSVRC Challenge 2014. It also achieves outstanding performance on the other image recognition datasets. Hence, we have also used VGG19 along with CovNet30 for generating sub-models. 

To fine-tune the VGG19 on X-ray images, the top layers (Fully-connected layer, and Softmax layer) of the VGG19 network are removed. We added new layers such as two Convolutional layers with ReLU activation, a Global Average Pooling layer, a Fully-connected layer, and a Softmax layer, at the top of the VGG19 network. Hyper parameters for the fine-tuning of VGG19 are same as the CovNet30. 

\subsection{Stacked Convolutional Neural network} 

Stacked generalization \cite{wolpert1992stacked} is an ensemble approach in which a new model learns how to incorporate the best predictions of multiple existing models. The proposed approach hypothesized that different CNN\textquotesingle s sub-models learn non-linear discriminative features and semantic image representation from the images at different levels. Thus a stacked ensemble CNN model will be generalized and highly accurate. This section describes the proposed stacked  convolutional neural network.  

 \begin{algorithm}[!hb]
 \caption{Sub-model Generation process}
 \begin{algorithmic}[1]
 \renewcommand{\algorithmicrequire}{\textbf{Input:}}
 \renewcommand{\algorithmicensure}{\textbf{Output:}}
 \REQUIRE X-ray images of the chest
 \ENSURE sub-models
 \\ 
  \STATE Divide Covid19CXr dataset into training set, validation set and test set.
  \STATE Apply data augmentation on train set. 
  \STATE Train CovNet30 and VGG19, and generate sub-models:  
 \\ \textit{Initialisation} : class\_weight = [0:30, 1:1, 1:1]
  \FOR {$i = 1$ to N}
  \STATE Train(CovNet30, train\_img, img\_label, class\_weight)
  \IF {($i == l1 $)}
  \STATE sub-model\#1 = save(CovNet30)
  \ENDIF
  \ENDFOR
  \STATE sub-model\#2 = save(CovNet30)
  \FOR {$i = 1$ to M}
  \STATE Train(VGG19, train\_img, img\_label, class\_weight)
  \IF {($i == l1$)}
  \STATE sub-model\#3 = save(VGG19)
  \ELSE \IF {($i == l2$)}
  \STATE sub-model\#4 = save(VGG19)
  \ENDIF
  \ENDIF
  \ENDFOR
  \STATE sub-model\#5 = save(VGG19)
  \RETURN \textit{sub-models} 
 \end{algorithmic} 
 \end{algorithm} 
 The pseudo-code of sub-models generation process is given in Algorithm 1. In this process the Covid19CXr dataset is divided in the train set, validation set and test set. The CovNet30 and VGG19 are trained on chest x-ray images of the training set for the 1530 iterations and 2121, respectively. During training of CovNet30, we have extracted the first \textit{sub-model\#1} after 765 iterations and second \textit{sub-model\#2} after completion of the training. 
Similarly, during fine-tuning of VGG19, we have extracted \textit{sub-model\#3} after 707 iterations, \textit{sub-model\#4} after 1414 iterations, and \textit{sub-model\#5} at last.

To deal with the class imbalance problem, we have assigned class weights while training of the networks. In this process, class weight in ratio of 30:1:1 is assigned to COVID-19, Pneumonia, and Normal class, respectively. 

The performance of the sub-models varies across complex CAD systems, and it is reasonable to combine the strengths of sub-models which might result in increased overall accuracy. Hence, we combined the sub-models output by logistic regression \cite{breiman1996stacked} to build a highly accurate and reliable generalized model. 
The pseudo-code of the process of creating stacked CNN is represented in Algorithm 2.

 \begin{algorithm} [!h]
 \caption{Stacked Convolutional Neural network and X-ray image Classification}
 \begin{algorithmic}[1]
 \renewcommand{\algorithmicrequire}{\textbf{Input:}}
 \renewcommand{\algorithmicensure}{\textbf{Output:}}
 \REQUIRE Validation set, test set, and sub-models
 \ENSURE Classification results
 \\ 
  \STATE Sub-Models Stacking: 
  \FOR {$i = 1$ to $length(validation\ \ set)$}
  \FOR {$j = 1$ to $5$}
  \STATE $[P1_{ji},\ \ P2_{ji},\ \ P3_{ji}]$ = sub-model\#(j).predict(validation\_img[i])
  \ENDFOR
  \STATE P = concatenation($[P1_{ji},\ \ P2_{ji},\ \ P3_{ji}]$
  \ENDFOR
  \STATE Fit logistic regression on feature vector P
  \STATE stacked\_model = fit.regression(P, validation\_label)
  \STATE Classifies X-ray image 
  \STATE pred\_label = classify(stacked\_model, test\_img) 
  \RETURN $pred\_label$ 
 \end{algorithmic} 
 \end{algorithm}   

The hypothesis representation of logistic regression model is as shown below:

\begin{equation} 
H_{\theta}^i(x)=g(\theta^{T}x)
\end{equation}

\begin{equation} 
where\   \ g(z)=\frac{1}{1+ e^{-z}}
\end{equation}

\begin{equation} 
hence\   \ H_{\theta}^i(x) =\frac{1}{1+ e^{-\theta^{T}x}}
\end{equation}

Where $H_{\theta}^i(x)$ is estimated probability $p(y=i/x; \theta)$ for each class $i$ (where $i \in \{0:COVID-19, 1:Normal, 2: Pneumonia\})$ of a image x. 

Next, we train a logistic regression model $H_{\theta}^i(x)$ using a one-vs-rest scheme \cite{scikit-learn} for each class $i$. For its training, we prepared dataset by providing X-ray images from the validation set to each of the sub-models and collects predictions.

In this case, every sub-model $j$ predicts three probabilities ($P1_{ji},\ \ P2_{ji},\ \ P3_{ji}$) for each X-ray image $i$ of that a given image $i$ belongs to each of the COVID-19, Normal, and Pneumonia classes. 
Let's say \textit{M} X-ray images in the validation set, and we concatenate the output probabilities of these five sub-models that become our feature vector $P_{M\times15}$  for the training of the logistic regression model. After training of the stacked model, on new input image \textit{x} from the test set, to make a prediction, pick the class $i$ that maximizes $max_i (H_{\theta}^i(x))$.

\section {EXPERIMENTS} 
This section presents the details of the dataset, evaluation metrics, experiments results and performance comparison.    

\subsection{Covid19CXr Dataset Generation}
In order to train and evaluate the performance of the proposed model, we generated X-ray images dataset, with the combination and modification of three publicly available datasets \cite{Chung2020covid, cohen2020covid,kermany2018labeled}, which are referred as COVID19CXr. 
\begin{figure}[!h]
        \centering
    \begin{subfigure}{.5\textwidth}
        \centering
        \includegraphics[width=.9\linewidth]{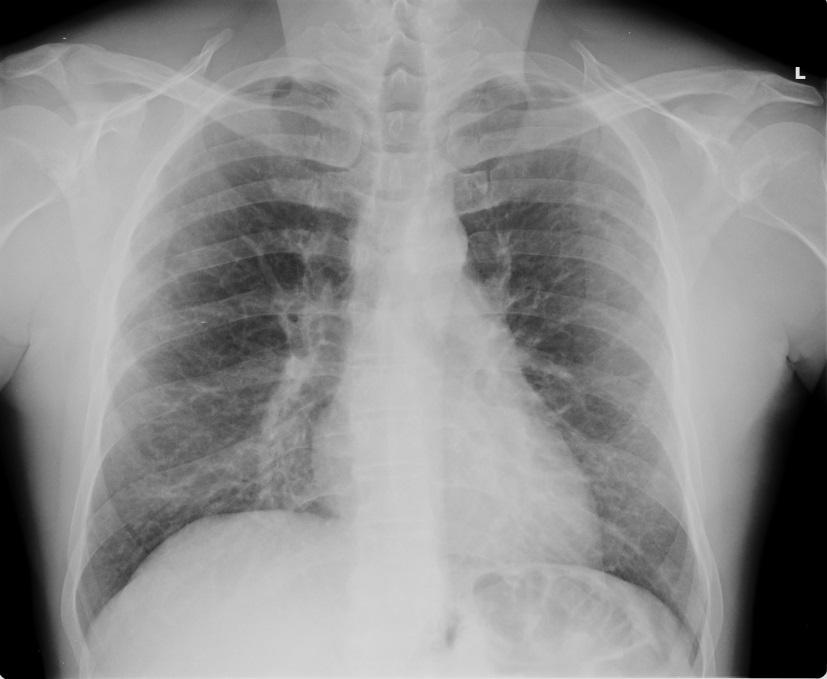}
        \caption{COVID-19}
        
    \end{subfigure}%
    \begin{subfigure}{.5\textwidth}
        \centering
        \includegraphics[width=.9\linewidth]{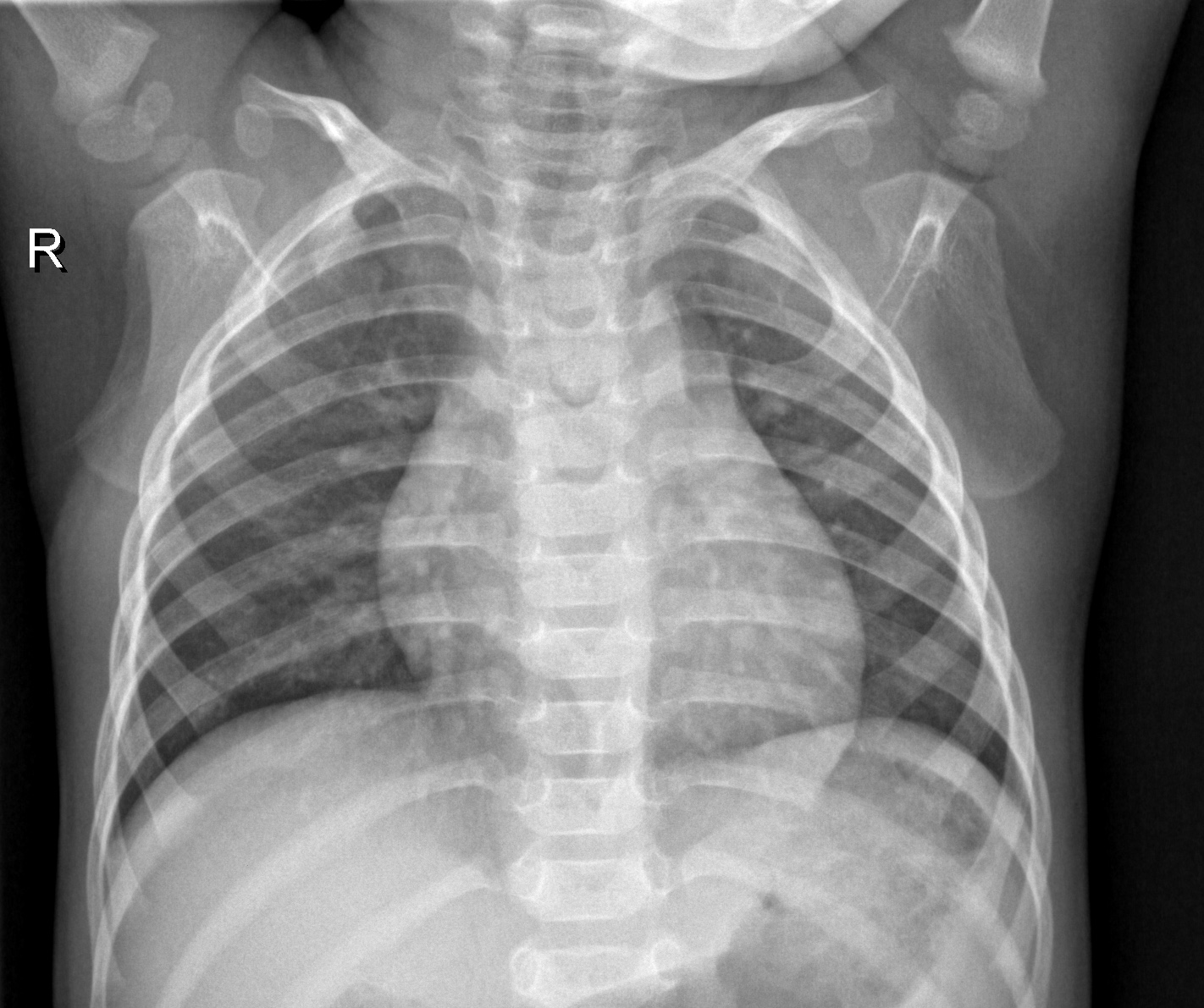}
        \caption{NORMAL}
    \end{subfigure}
    \begin{subfigure}{.5\textwidth}
        \centering
        \includegraphics[width=.9\linewidth]{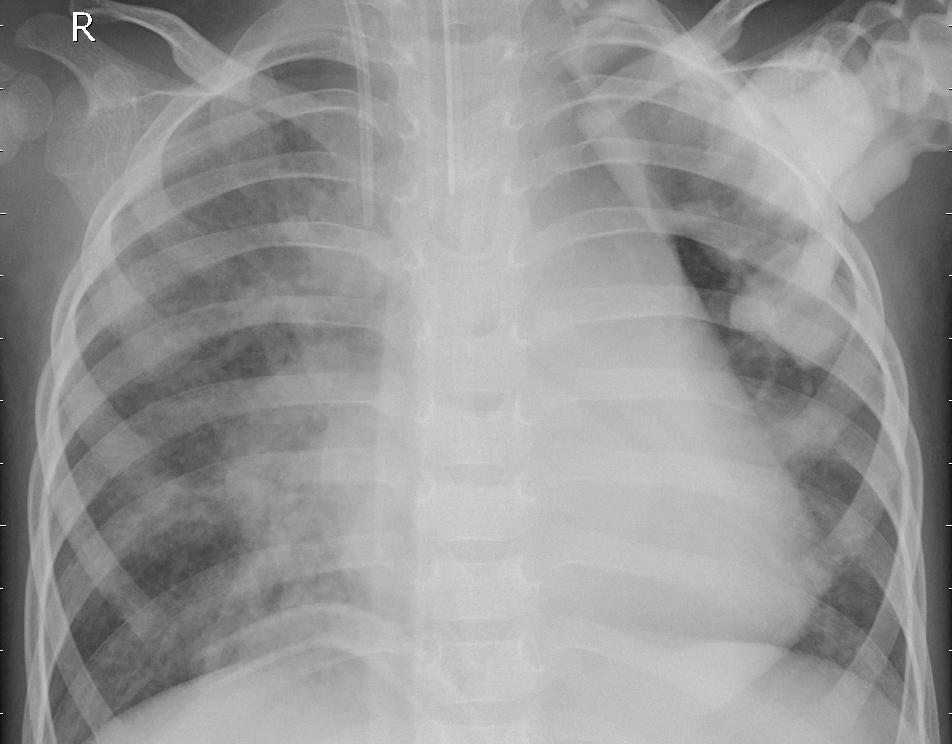}
        \caption{PNEUMONIA}
        \end{subfigure}
\caption{Samples of X-ray images of Chest from the database; where image in (a) COVID-19, (b) NORMAL, (c) PEUMONIA}
    \label{fig:sampleimage1}
\end{figure}
The COVID19CXr includes 2764 chest X-ray images of 1768 patients. Out of 2764 images, 270 images of 170 patients belong to a COVID-19 class, 1139 images of 1015 patients belong to Normal class and 1355 images of 583 patients belong to a Pneumonia class. COVID-19 images are obtained from the two publicly available repositories: 1) \textit{``Figure-1 COVID-19 Chest X-ray Dataset Initiative''} \cite{Chung2020covid} and 2) \textit{``COVID-19 Image Data Collection''} \cite{cohen2020covid}. Pneumonia and Normal cases images are included from the \textit{``Mendeley data 2''} \cite{kermany2018labeled}. Figure \ref{fig:sampleimage1} 
shows the sample chest X-ray images of COVID-19, Normal and Pneumonia classes from the COVID19CXr dataset. 

\begin{table}[!h]
\caption{Images distribution in the train set, validation set and test set, corresponding to  folds}
\label{tab:folds}
\begin{adjustbox}{max width=\textwidth}
\begin{tabular}{clcccc} \hline
{ \textbf{Fold (s)}} &
  \multicolumn{1}{c}{{\textbf{Data set (s)}}} &
  {\textbf{COVID-19}} &
  {\textbf{Normal}} &
  {\textbf{Pneumonia}} &
  {\textbf{Total}} \\ \hline
{ } &
  { \textbf{Train Set}} &
  { 189} &
  { 798} &
  { 948} &
  { \textbf{1935}} \\ 
{ } &
  { \textbf{Validation   set}} &
  { 25} &
  { 113} &
  { 136} &
  { \textbf{274}} \\
\multirow{-3}{*}{{ Fold1}} &
  { \textbf{Test Set}} &
  { 56} &
  { 228} &
  { 271} &
  { \textbf{555}} \\ \cline{1-6}
{ } &
  { \textbf{Train Set}} &
  { 190} &
  { 799} &
  { 951} &
  { \textbf{1940}} \\
{ } &
  { \textbf{Validation   set}} &
  { 27} &
  { 113} &
  { 135} &
  { \textbf{275}} \\
\multirow{-3}{*}{{ Fold2}} &
  { \textbf{Test Set}} &
  { 53} &
  { 227} &
  { 269} &
  { \textbf{549}} \\ \cline{1-6}
{ } &
  { \textbf{Train Set}} &
  { 189} &
  { 798} &
  { 950} &
  { \textbf{1937}} \\
{ } &
  { \textbf{Validation   set}} &
  { 26} &
  { 113} &
  { 134} &
  { \textbf{273}} \\
\multirow{-3}{*}{{ Fold3}} &
  { \textbf{Test Set}} &
  { 55} &
  { 228} &
  { 271} &
  { \textbf{554}} \\ \cline{1-6}
{ } &
  { \textbf{Train Set}} &
  { 193} &
  { 797} &
  { 945} &
  { \textbf{1935}} \\
{ } &
  { \textbf{Validation   set}} &
  { 27} &
  { 114} &
  { 136} &
  { \textbf{277}} \\
\multirow{-3}{*}{{ Fold4}} &
  { \textbf{Test Set}} &
  { 50} &
  { 228} &
  { 274} &
  { \textbf{552}} \\ \cline{1-6}
{ } &
  { \textbf{Train Set}} &
  { 189} &
  { 798} &
  { 952} &
  { \textbf{1939}} \\
{ } &
  { \textbf{Validation   set}} &
  { 25} &
  { 113} &
  { 136} &
  { \textbf{274}} \\
\multirow{-3}{*}{{ Fold5}} &
  { \textbf{Test Set}} &
  { 56} &
  { 228} &
  { 267} &
  { \textbf{551}} \\ \hline

\end{tabular}
\end{adjustbox}

\end{table}
For the performance assessment of the proposed method, we have used 5-fold cross-validation approach, in which the dataset is divided into 5-folds at the patient level. Table \ref{tab:folds} gives details of the distribution of images in the training set, validation set, and test set corresponding to each fold. The training set and validation set are used while training the network, and an hold-out test set is used for the performance assessment of the proposed model.   

\subsection{Evaluation Metrics}
To assess the performance of proposed method we have used sensitivity,	specificity, accuracy, positive prediction value (PPV), F1-score, area under the curve (AUC) and confusion matrix as evaluation metrics. The mathematical definition for the evaluation metrics is given below (in Eqn. (5), Eqn. (6), Eqn. (7), Eqn. (8) and Eqn. (9) respectively): 

\begin{equation}
Accuracy = \frac{(TP+TN)}{(TP+TN+FP+FN)} \end{equation}

\begin{equation}
PPV =  \frac{TP}{(TP+FP)}
\end{equation}  
\begin{equation}
Sensitivity = \frac{TP}{(TP+FN)}
\end{equation}

\begin{equation}
Specificity = \frac{TN}{(TN+FP)}
\end{equation}
\begin{equation}
F1-Score =  \frac{2TP}{(2TP+FP+FN)}
\end{equation} 
Where true positive (TP), true negative (TN), false positive (FP), and false negative (FN) are the parameters of confusion matrix. The present study deals with a multi-class problem; therefore, to get the overall metric score of the method, we calculated the mean of each metric. 
\begin{table}[!b]
\caption{Diagnosis performance of stacked CNN model.}
\begin{adjustbox}{max width=\textwidth}
\begin{tabular}{cccccccc} \hline
\textbf{Fold (s)} &
  \textbf{\begin{tabular}[c]{@{}c@{}}Sensitivity \\ (\%)\end{tabular}} &
  \textbf{\begin{tabular}[c]{@{}c@{}}Specificity \\ (\%)\end{tabular}} &
  \textbf{\begin{tabular}[c]{@{}c@{}}Accuracy\\ (\%)\end{tabular}} & \textbf{\begin{tabular}[c]{@{}c@{}}Err $ \pm $ CI \\  (\%) \end{tabular}} &
  \textbf{\begin{tabular}[c]{@{}c@{}}PPV\\ (\%)\end{tabular}} &
  \textbf{F1-Score} &
  \textbf{AUC $ \pm $ CI} \\ \hline
Fold1         & 95.5           & 98.19          & 96.94       & 3.06$ \pm 1.42$  & 97.69          & 0.97          & 0.989$ \pm 0.003$          \\
Fold2         & 92.66          & 94.97          & 91.44        & 8.56$ \pm 2.39$ & 91.32          & 0.92          & 0.982$ \pm 0.015$          \\
Fold3         & 91.45          & 95.01          & 91.34        & 8.66$ \pm 2.35$  & 92.13          & 0.92          & 0.982$ \pm 0.011$          \\
Fold4         & 92.47          & 94.77          & 90.22         & 9.78$\pm 2.48$ & 85.97          & 0.88          & 0.977$ \pm 0.023$          \\
Fold5         & 94.59          & 96.12          & 93.74         & 6.26$ \pm 2.02$ & 93.54          & 0.94          & 0.981$ \pm 0.009$          \\
\textbf{Mean} & \textbf{93.33} & \textbf{95.81} & \textbf{92.74} & \textbf{7.26}$ \pm \textbf{2.16} $ &\textbf{92.13} & \textbf{0.93} & \textbf{0.984}$ \pm \textbf{0.012}$ \\ \hline 
\multicolumn{6}{c} {\textit{``Err $ \pm $ CI": classification error (Err) with  95\% confidence interval (CI).}} 
\\
\multicolumn{7}{l}{\textit{``AUC $ \pm $ CI": area under the curve (AUC) with  95\% confidence interval (CI).}}
\end{tabular}
\end{adjustbox}
\end{table}
\subsection{Results and discussion}

In order to evaluate the performance of our proposed stacked  convolutional neural network, we conduct a set of experiments. 
In the first experiment, data augmentation techniques such as flip, rotation, shear, zoom, and shift have been applied on a training set. Thereafter, the augmented training set is utilized for the training of CovNet30 model and fine-tuning VGG19 model. In the second experiment, the stacked CNN model is trained on the validation set. Finally, evaluation results are produced on the test set. We repeat the same set of experiments five times for each fold. The following sections represent the experimental results and performance comparison. 

\begin{figure}[!b]
        \centering
    \begin{subfigure}{.5\textwidth}
        \centering
        \includegraphics[width=1.1\linewidth]{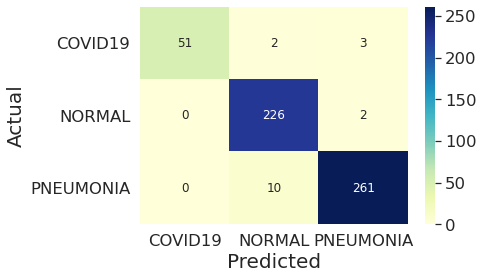}
        \caption{Fold1}
      
    \end{subfigure}%
    \begin{subfigure}{.5\textwidth}
        \centering
        \includegraphics[width=1.1\linewidth]{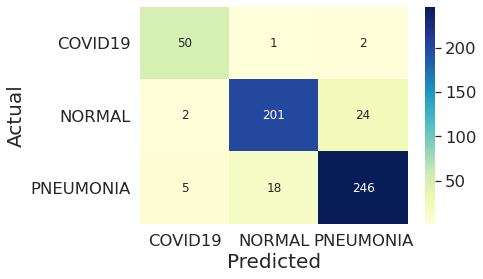}
        \caption{Fold2}
     
    \end{subfigure}
    \begin{subfigure}{.5\textwidth}
        \centering
        \includegraphics[width=1.1\linewidth]{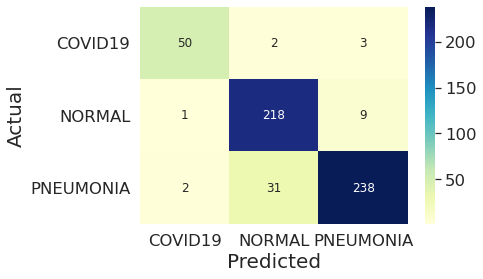}
        \caption{Fold3}
        \end{subfigure}%
     \begin{subfigure}{.5\textwidth}
        \centering
        \includegraphics[width=1.1\linewidth]{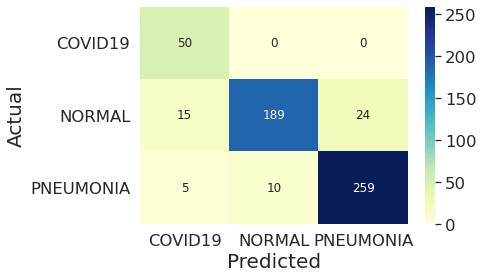}
        \caption{Fold4}
     \end{subfigure}
    \begin{subfigure}{.5\textwidth}
        \centering
        \includegraphics[width=1.2\linewidth]{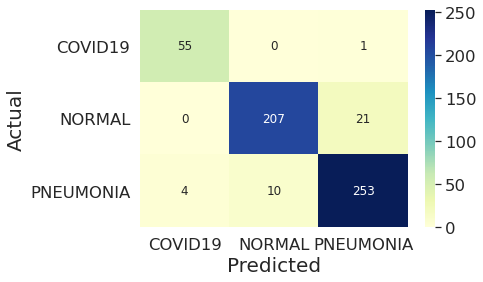}
        \caption{Fold5}
        \end{subfigure}

\caption{Confusion matrix for stacked CNN model on the different folds.}
    \label{fig:sampleimage3}
\end{figure}
\begin{figure}[bh!]
        \centering
    \begin{subfigure}{.5\textwidth}
        \centering
        \includegraphics[width=.99\linewidth]{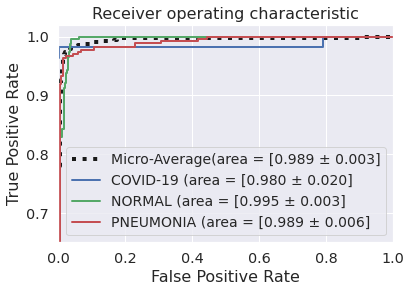}
        \caption{Fold1}
      
    \end{subfigure}%
    \begin{subfigure}{.5\textwidth}
        \centering
        \includegraphics[width=.99\linewidth]{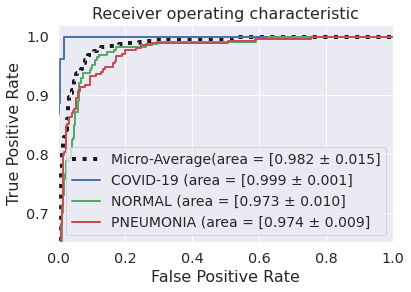}
        \caption{Fold2}
     
    \end{subfigure}
    \begin{subfigure}{.5\textwidth}
        \centering
        \includegraphics[width=.99\linewidth]{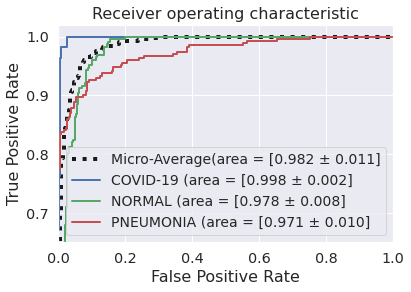}
        \caption{Fold3}
        \end{subfigure}%
     \begin{subfigure}{.5\textwidth}
        \centering
        \includegraphics[width=.99\linewidth]{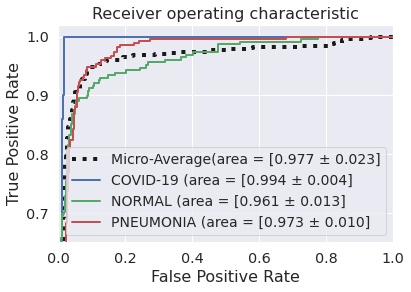}
        \caption{Fold4}
     \end{subfigure}
    \begin{subfigure}{.5\textwidth}
        \centering
        \includegraphics[width=.99\linewidth]{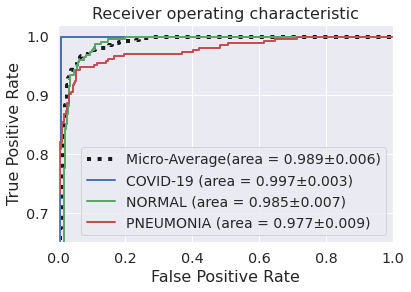}
        \caption{Fold5}
        \end{subfigure}
\caption{ROC curve for stacked CNN model on the different folds.}
    \label{fig:sampleimage4}
\end{figure}

\subsubsection{Discrimination power of stacked CNN model:} Table 3 presents the diagnostic performance of stacked  CNN and it shows good discrimination ability for the diagnosis of the COVID-19 from the chest X-ray images. The proposed model achieved mean sensitivity of 93.33 \%, specificity of 95.81 \%, PPV of 92.74 \%, and  accuracy of 92.74 \% to classify: COVID-19, Normal and Pneumonia classes. Our method achieved good sensitivity; so that we can limit the miss classification of the COVID-19 positive cases. In addition, the confidence interval for the classification error and AUC is calculated at the 95\% confidence level. As shown in Table 3 the classification error of the proposed model is  \textit{$7.26 \% \pm2.16\% $} at the 95\% confidence level.

For a deeper exploration of the performances of the proposed method the confusion matrix and receiver operating characteristic (ROC) curve (with AUC\textquotesingle s CI) corresponding to each fold are evaluated and shown in Figure 3 and Figure 4, respectively. It can be observed from the confusion matrix the proposed model produce very less false negative and false positive, specifically for the COVID-19 cases compared to other cases of COVID19CXr dataset. For COVID-19 cases, it is essential to minimize the wrong diagnosis. On the other hand, the ROC curve shows the stability of the proposed stacked  CNN model, and the present model achieved a mean AUC of 0.994 for COVID-19 class and a mean AUC of 0.984 along with CI of [0.012]for all categories.    

\subsubsection{Performance comparison:} 
Table 4 shows the performance comparison of the proposed method, VGG19 and CovNet30. It is observed from Table 4 that the proposed model achieves the best accuracy of 92.74 \% compared to stand-alone model VGG19 and CovNet30, our stacked CNN model improves the diagnosis accuracy by $2.88 \% \sim 5.01\%$. 
\begin{table}[b!]
\caption{Performance comparison for different methods.}
\centering
\begin{tabular}{lccc} \hline
\multicolumn{1}{c}{\multirow{2}{*}{\textbf{Metric}}} & \multicolumn{3}{c}{\textbf{Method}} \\ \cline{2-4}
\multicolumn{1}{c}{}          & VGG19      & CovNet30   & Proposed    \\ \hline
\textbf{Accuracy (\%)}                     & 89.86${\pm}$2.21     & 87.73${\pm}$3.08 & 92.74${\pm}$2.68 \\
\textbf{Sensitivity (\%)}    & 90.34${\pm}$4.24 & 86.80${\pm}$4.05 & 93.33${\pm}$1.66 \\
\textbf{Specificity (\%)}    & 94.15${\pm}$1.26 & 93.49${\pm}$1.39 & 95.81${\pm}$1.43 \\
\textbf{PPV (\%)}             & 90.53${\pm}$2.96 & 84.54${\pm}$5.23 & 92.13${\pm}$4.23 \\
\textbf{F1-score}                & 0.90${\pm}$0.03  & 0.85${\pm}$0.04  & 0.93${\pm}$0.03  \\
\textbf{AUC}                     & 0.97${\pm}$0.01  & 0.97${\pm}$0.02  & 0.98${\pm}$0.01 \\ \hline
\end{tabular}
\end{table}
In terms of sensitivity, specificity, and PPV, proposed model also shows significant performance improvements by $2.99\% \sim 6.53\%$, $1.66\% \sim 2.32\%$, and $1.6\% \sim 7.59\%$,  respectively. 
\begin{table}[t!]
\caption{Performance evaluation of the sub-models.}
\centering
\begin{tabular}{lcccccc} \hline
\multicolumn{2}{c}{\textbf{Model (s)}}     & \textbf{Fold1} & \textbf{Fold2} & \textbf{Fold3} & \textbf{Fold4} & \textbf{Fold5} \\ \hline
\multirow{2}{*}{\textbf{CovNet30}} & Sub-model1 & 86.3 & 89.4 & 51.6 & 59.1 & 84   \\
                                   & Sub-model2 & 95.7 & 91.1 & 86.5 & 83.5 & 90.4 \\ \cline{1-7}
\multirow{3}{*}{\textbf{VGG19}}    & Sub-model3 & 85.4 & 84.9 & 89.4 & 85.9 & 57.5 \\
                                   & Sub-model4 & 93.3 & 87.8 & 79.6 & 90.2 & 79.9 \\
                                   & Sub-model5 & 92.8 & 89.4 & 87.9 & 90.4 & 91.1 \\ \cline{1-7}
\multicolumn{2}{c}{\textbf{Proposed model}} & \textbf{96.94} & \textbf{91.44} & \textbf{91.34} & \textbf{90.22} & \textbf{93.54}\\ \hline
\end{tabular} 
\end{table}
\begin{table}[!ht]
\caption{Performance comparison with existing methods}
\begin{adjustbox}{max width=\textwidth}
\begin{tabular}{lccccc} \hline
\multicolumn{1}{c}{\multirow{2}{*}{\textbf{Author (s)}}} &
  \multirow{2}{*}{\textbf{Method}} &
  \multicolumn{2}{c}{\textbf{Dataset}} &
  \multirow{2}{*}{\textbf{Classification Task}} &
  \multirow{2}{*}{\textbf{Results}} \\ \cline{3-4}
\multicolumn{1}{c}{} &
   &
 \textbf{ Modality} &
  \textbf{Subjects} &
   &
   \\ \hline
Narin et al. \cite{narin2020automatic} &
  Pre-Train   CNN &
  X-ray &
  \begin{tabular}[c]{@{}c@{}}50 COVID-19,\\ 50 Normal\end{tabular} &
   \begin{tabular}[c]{@{}c@{}} Binary:  COVID-19,\\ Normal \end{tabular} &
  97 \%   (Acc) \\ \hline
Wang et al. \cite{wang2020covid} &
  COVID-Net &
  X-ray &
  \begin{tabular}[c]{@{}c@{}}183 COVID-19,\\ 8066 Normal, \\ 5538 non-COVID19\end{tabular} &
  \begin{tabular}[c]{@{}c@{}}Multiclass:  COVID-19,\\ Normal, Non-COVID19\end{tabular} &
  92.6\%   (Acc) \\ \hline
Oh et al. \cite{oh2020deep} &
  ResNet18 &
  X-ray &
  \begin{tabular}[c]{@{}c@{}}191 Normal, \\54 Bacterial, \\ 57 Tuberculosis,\\20  Viral, \\ 180 COVID-19\end{tabular} &
  \begin{tabular}[c]{@{}c@{}} Multiclass:  Normal,\\ Bacterial, Tuberculosis, \\ Viral,   COVID-19\end{tabular} &
  88.9\%   (Acc) \\ \hline
\multirow{2}{*}{Ozturk et al. \cite{ozturk2020automated}} &
  \multirow{2}{*}{DarkCovidNet} &
  \multirow{2}{*}{X-ray} &
  \multirow{2}{*}{\begin{tabular}[c]{@{}c@{}}127 COVID-19,\\   500 no-findings, \\ 500 pneumonia\end{tabular}} &
  \begin{tabular}[c]{@{}c@{}} Binary: COVID-19, \\ No-findings\end{tabular} &
  98.08\% (Acc) \\ \cline{5-6}
 &
   &
   &
   &
  \begin{tabular}[c]{@{}c@{}}Multiclass: COVID-19, \\ No-findings, Pneumonia\end{tabular} &
  87.02\%   (Acc) \\ \hline
Pereira et al.    \cite{pereira2020covid} &
  \begin{tabular}[c]{@{}c@{}}Deep features,\\ Texture features, \\Fusion techniques\end{tabular} &
  X-ray &
  \begin{tabular}[c]{@{}c@{}}200 Normal, \\180 COVID-19,\\ 22 SARS, \\20 MERS, \\ 22 Pneumocystis, \\ 24 Streptococcus, \\20 Varicella\end{tabular} &
  \begin{tabular}[c]{@{}c@{}}Hierarchical: Normal, \\ COVID-19, SARS,\\ MERS, Pneumocystis, \\ Streptococcus, Varicella\end{tabular} &
  0.89   (F1-score) \\ \hline
Ucar et al. \cite{ucar2020covidiagnosis} &
  SqueezeNet CNN &
  X-ray &
  \begin{tabular}[c]{@{}c@{}}1583 Normal, \\ 4290 Pneumonia, \\ 76 COVID-19\end{tabular} &
  \begin{tabular}[c]{@{}c@{}}Multiclass:  Normal, \\ Pneumonia, COVID-19\end{tabular} &
  \begin{tabular}[c]{@{}c@{}}95.7 \% (Acc),\\ 90 \% (Sens)\end{tabular} \\ \hline
Sethy et al. \cite{sethy2020detection} &
  Deep feature, SVM &
  X-ray &
  \begin{tabular}[c]{@{}c@{}}25 COVID-19+,\\ 25 COVID-19-\end{tabular} &
  \begin{tabular}[c]{@{}c@{}}Binary: COVID-19+, \\ COVID-19-\end{tabular} &
  95.38\% (Acc) \\ \hline
 
\textbf{Proposed Method} &
  Stacked  CNN &
  X-ray &
  \begin{tabular}[c]{@{}c@{}}270 COVID-19,\\ 1139 Normal, \\ 1355 Pneumonia \end{tabular} &
  \begin{tabular}[c]{@{}c@{}}Multiclass: COVID-19, \\ Normal, Pneumonia \end{tabular} &
  \begin{tabular}[c]{@{}c@{}} 92.74\% (Acc) \\ 93.33 \% (Sens) \\ 0.93 (F1-Score)\end{tabular} \\  \hline 
\end{tabular}
\end{adjustbox}
\end{table}
Further, Table 5 represents the performance of the individual sub-models and the proposed stacked CNN model corresponding to each fold. Our stacked ensemble CNN model achieved better performance as compared to all sub-models, over the each fold.  

A variety of deep learning-based studies have already been proposed in past studies for COVID-19 diagnosis from the chest X-ray images. The performance comparison of the proposed method in the present study with some of related studies are shown in the Table 6.

Since COVID-19 is a new epidemic and there are limited number of COVID-19 X-ray images are  available publicly for developing CAD systems for COVID-19 detection. Studies in \cite{sethy2020detection} and  \cite{narin2020automatic} have just developed on the 25 and 50 images for each class, respectively. Other studies in the \cite{wang2020covid, oh2020deep, ozturk2020automated, pereira2020covid, ucar2020covidiagnosis} have used less 200 COVID-19 images for developing their methods. In this study, a total of 2764 X-ray images has been used to develop our model, including  270 COVID-19 images, which the largest number of COVID-19 images among all the studies in Table 6.
It can be observed from Table 6 that for the multi-class classification task, the proposed approach shows the superiority over the methods in \cite{wang2020covid, oh2020deep, ozturk2020automated, pereira2020covid}, except the method in \cite{ucar2020covidiagnosis}, which has higher accuracy. However, sensitivity is higher for our approach.    

Some of the salient features of stacked  CNN can be summarized as:
\begin{itemize}
  \item The proposed method is based on the stacked generalization of CNN\textquotesingle s sub-models, which minimizes the variance of predictions and reduces generalization error. As of the result,  stacked CNN yields high diagnosis accuracy in the X-ray images.  
  
  \item The proposed stacked  CNN model produces very less false positive (type 1) and false negative (type  2) error, which confirms that the stacked  CNN is reliable for clinical uses.
  \item The proposed model is developed based on a less complex network, which computationally efficient and shows its stability on a small dataset.  
\end{itemize}

\section{CONCLUSION}
In this paper, we introduced a new stacked convolutional neural network for the automatic diagnosis of the COVID19 from the Chest X-ray images. In the proposed method, CNN\textquotesingle s sub-models have obtained from our developed CovNet30 model and pre-trained VGG19 model. Stacked CNN model ensemble the sub-models using logistic regression, for deriving a powerful model for image classification than individual sub-models.
The proposed model is able to learn image discriminative features and retrieved the diverse information present in the X-ray images of the chest. 
It achieves a classification accuracy of 92.74\%, sensitivity of 93.33\%, PPV of 92.13\%, and F1-score of 0.93 on the chest X-ray images of COVID19CXr dataset. Our proposed approach shows its superiority over the existing methods for the diagnosis of the COVID-19 from the X-ray images. 

Our experiments results show the effectiveness of the stacked  CNN for classification of COVID-19, Normal, and Pneumonia X-ray images. More importantly, the proposed model outperforms the pre-trained VGG19 and CovNet30 model for the classification of X-ray images.

% Since the last 40 years, many computer-aided systems have been developed for the diagnosis of lung disease \cite{doi2007computer} and achieved promising results for automatic detecting lung abnormality from the radiological images \cite{castellano2004texture, van2002automatic,jaeger2013automatic}. Recently, many researchers  

%\section*{References}

%\bibliography{mybibfile}

\begin{thebibliography}{10}
\expandafter\ifx\csname url\endcsname\relax
  \def\url#1{\texttt{#1}}\fi
\expandafter\ifx\csname urlprefix\endcsname\relax\def\urlprefix{URL }\fi
\expandafter\ifx\csname href\endcsname\relax
  \def\href#1#2{#2} \def\path#1{#1}\fi

\bibitem{nishiura2020serial}
H.~Nishiura, N.~M. Linton, A.~R. Akhmetzhanov, Serial interval of novel
  coronavirus (covid-19) infections, International journal of infectious
  diseases (2020).

\bibitem{chen2020emerging}
Y.~Chen, Q.~Liu, D.~Guo, Emerging coronaviruses: genome structure, replication,
  and pathogenesis, Journal of medical virology 92~(4) (2020) 418--423.

\bibitem{world2020novel}
W.~H. Organization, et~al., Novel coronavirus (2019-ncov) technical guidance:
  laboratory testing for 2019-ncov in humans, World Health Organization,
  Geneva, Switzerland. https://www. who.
  int/emergencies/diseases/novel-coronavirus-2019/technical-guidance/laboratory-guidance
  (2020).

\bibitem{ng2020imaging}
M.-Y. Ng, E.~Y. Lee, J.~Yang, F.~Yang, X.~Li, H.~Wang, M.~M.-s. Lui, C.~S.-Y.
  Lo, B.~Leung, P.-L. Khong, et~al., Imaging profile of the covid-19 infection:
  radiologic findings and literature review, Radiology: Cardiothoracic Imaging
  2~(1) (2020) e200034.

\bibitem{huang2020clinical}
C.~Huang, Y.~Wang, X.~Li, L.~Ren, J.~Zhao, Y.~Hu, L.~Zhang, G.~Fan, J.~Xu,
  X.~Gu, et~al., Clinical features of patients infected with 2019 novel
  coronavirus in wuhan, china, The lancet 395~(10223) (2020) 497--506.

\bibitem{Chung2020covid}
C.~et~al.,
  \href{https://github.com/agchung/Figure1-COVID-chestxraydataset}{Figure 1
  covid-19 chest x-ray data initiative.} (2020).
\newline\urlprefix\url{https://github.com/agchung/Figure1-COVID-chestxraydataset}

\bibitem{cohen2020covid}
J.~P. Cohen, P.~Morrison, L.~Dao,
  \href{https://github.com/ieee8023/covid-chestxray-dataset}{Covid-19 image
  data collection}, arXiv preprint arXiv:2003.11597 (2020).
\newline\urlprefix\url{https://github.com/ieee8023/covid-chestxray-dataset}

\bibitem{kermany2018labeled}
D.~Kermany, K.~Zhang, M.~Goldbaum, Labeled optical coherence tomography (oct)
  and chest x-ray images for classification, Mendeley data 2 (2018).

\bibitem{doi2007computer}
K.~Doi, Computer-aided diagnosis in medical imaging: historical review, current
  status and future potential, Computerized medical imaging and graphics
  31~(4-5) (2007) 198--211.

\bibitem{castellano2004texture}
G.~Castellano, L.~Bonilha, L.~Li, F.~Cendes, Texture analysis of medical
  images, Clinical radiology 59~(12) (2004) 1061--1069.

\bibitem{van2002automatic}
B.~Van~Ginneken, S.~Katsuragawa, B.~M. ter Haar~Romeny, K.~Doi, M.~A.
  Viergever, Automatic detection of abnormalities in chest radiographs using
  local texture analysis, IEEE transactions on medical imaging 21~(2) (2002)
  139--149.

\bibitem{jaeger2013automatic}
S.~Jaeger, A.~Karargyris, S.~Candemir, L.~Folio, J.~Siegelman, F.~Callaghan,
  Z.~Xue, K.~Palaniappan, R.~K. Singh, S.~Antani, et~al., Automatic
  tuberculosis screening using chest radiographs, IEEE transactions on medical
  imaging 33~(2) (2013) 233--245.

\bibitem{shi2020review}
F.~Shi, J.~Wang, J.~Shi, Z.~Wu, Q.~Wang, Z.~Tang, K.~He, Y.~Shi, D.~Shen,
  Review of artificial intelligence techniques in imaging data acquisition,
  segmentation and diagnosis for covid-19, IEEE Reviews in Biomedical
  Engineering (2020).

\bibitem{dong2020role}
D.~Dong, Z.~Tang, S.~Wang, H.~Hui, L.~Gong, Y.~Lu, Z.~Xue, H.~Liao, F.~Chen,
  F.~Yang, et~al., The role of imaging in the detection and management of
  covid-19: a review, IEEE Reviews in Biomedical Engineering (2020).

\bibitem{butt2020deep}
C.~Butt, J.~Gill, D.~Chun, B.~A. Babu, Deep learning system to screen
  coronavirus disease 2019 pneumonia, Applied Intelligence (2020) 1.

\bibitem{ardakani2020application}
A.~A. Ardakani, A.~R. Kanafi, U.~R. Acharya, N.~Khadem, A.~Mohammadi,
  Application of deep learning technique to manage covid-19 in routine clinical
  practice using ct images: Results of 10 convolutional neural networks,
  Computers in Biology and Medicine (2020) 103795.

\bibitem{narin2020automatic}
A.~Narin, C.~Kaya, Z.~Pamuk, Automatic detection of coronavirus disease
  (covid-19) using x-ray images and deep convolutional neural networks, arXiv
  preprint arXiv:2003.10849 (2020).

\bibitem{wang2020covid}
L.~Wang, A.~Wong, Covid-net: A tailored deep convolutional neural network
  design for detection of covid-19 cases from chest radiography images, arXiv
  preprint arXiv:2003.09871 (2020).

\bibitem{oh2020deep}
Y.~Oh, S.~Park, J.~C. Ye, Deep learning covid-19 features on cxr using limited
  training data sets, IEEE Transactions on Medical Imaging (2020).

\bibitem{ozturk2020automated}
T.~Ozturk, M.~Talo, E.~A. Yildirim, U.~B. Baloglu, O.~Yildirim, U.~R. Acharya,
  Automated detection of covid-19 cases using deep neural networks with x-ray
  images, Computers in Biology and Medicine (2020) 103792.

\bibitem{pereira2020covid}
R.~M. Pereira, D.~Bertolini, L.~O. Teixeira, C.~N. Silla~Jr, Y.~M. Costa,
  Covid-19 identification in chest x-ray images on flat and hierarchical
  classification scenarios, Computer Methods and Programs in Biomedicine (2020)
  105532.

\bibitem{wang2020prior}
J.~Wang, Y.~Bao, Y.~Wen, H.~Lu, H.~Luo, Y.~Xiang, X.~Li, C.~Liu, D.~Qian,
  Prior-attention residual learning for more discriminative covid-19 screening
  in ct images, IEEE Transactions on Medical Imaging (2020).

\bibitem{hu2020weakly}
S.~Hu, Y.~Gao, Z.~Niu, Y.~Jiang, L.~Li, X.~Xiao, M.~Wang, E.~F. Fang,
  W.~Menpes-Smith, J.~Xia, et~al., Weakly supervised deep learning for covid-19
  infection detection and classification from ct images, arXiv preprint
  arXiv:2004.06689 (2020).

\bibitem{ucar2020covidiagnosis}
F.~Ucar, D.~Korkmaz, Covidiagnosis-net: Deep bayes-squeezenet based diagnostic
  of the coronavirus disease 2019 (covid-19) from x-ray images, Medical
  Hypotheses (2020) 109761.

\bibitem{afshar2020covid}
P.~Afshar, S.~Heidarian, F.~Naderkhani, A.~Oikonomou, K.~N. Plataniotis,
  A.~Mohammadi, Covid-caps: A capsule network-based framework for
  identification of covid-19 cases from x-ray images, arXiv preprint
  arXiv:2004.02696 (2020).

\bibitem{sethy2020detection}
P.~K. Sethy, S.~K. Behera, Detection of coronavirus disease (covid-19) based on
  deep features, Preprints 2020030300 (2020) 2020.

\bibitem{krizhevsky2012imagenet}
A.~Krizhevsky, I.~Sutskever, G.~E. Hinton, Imagenet classification with deep
  convolutional neural networks, in: Advances in neural information processing
  systems, 2012, pp. 1097--1105.

\bibitem{he2016deep}
K.~He, X.~Zhang, S.~Ren, J.~Sun, Deep residual learning for image recognition,
  in: Proceedings of the IEEE conference on computer vision and pattern
  recognition, 2016, pp. 770--778.

\bibitem{simonyan2014very}
K.~Simonyan, A.~Zisserman, Very deep convolutional networks for large-scale
  image recognition, arXiv preprint arXiv:1409.1556 (2014).

\bibitem{girshick2014rich}
R.~Girshick, J.~Donahue, T.~Darrell, J.~Malik, Rich feature hierarchies for
  accurate object detection and semantic segmentation, in: Proceedings of the
  IEEE conference on computer vision and pattern recognition, 2014, pp.
  580--587.

\bibitem{girshick2015fast}
R.~Girshick, Fast r-cnn, in: Proceedings of the IEEE international conference
  on computer vision, 2015, pp. 1440--1448.

\bibitem{gour2019deeprnnetseg}
M.~Gour, S.~Jain, R.~Agrawal, Deeprnnetseg: Deep residual neural network for
  nuclei segmentation on breast cancer histopathological images, in:
  International Conference on Computer Vision and Image Processing, Springer,
  2019, pp. 243--253.

\bibitem{ronneberger2015u}
O.~Ronneberger, P.~Fischer, T.~Brox, U-net: Convolutional networks for
  biomedical image segmentation, in: International Conference on Medical image
  computing and computer-assisted intervention, Springer, 2015, pp. 234--241.

\bibitem{gour2020residual}
M.~Gour, S.~Jain, T.~S. Kumar, Residual learning based cnn for breast cancer
  histopathological image classification, International Journal of Imaging
  Systems and Technology (2020).

\bibitem{anwar2018medical}
S.~M. Anwar, M.~Majid, A.~Qayyum, M.~Awais, M.~Alnowami, M.~K. Khan, Medical
  image analysis using convolutional neural networks: a review, Journal of
  medical systems 42~(11) (2018) 226.

\bibitem{shen2017deep}
D.~Shen, G.~Wu, H.-I. Suk, Deep learning in medical image analysis, Annual
  review of biomedical engineering 19 (2017) 221--248.

\bibitem{huang2017lung}
X.~Huang, J.~Shan, V.~Vaidya, Lung nodule detection in ct using 3d
  convolutional neural networks, in: 2017 IEEE 14th International Symposium on
  Biomedical Imaging (ISBI 2017), IEEE, 2017, pp. 379--383.

\bibitem{rajpurkar2017chexnet}
P.~Rajpurkar, J.~Irvin, K.~Zhu, B.~Yang, H.~Mehta, T.~Duan, D.~Ding, A.~Bagul,
  C.~Langlotz, K.~Shpanskaya, et~al., Chexnet: Radiologist-level pneumonia
  detection on chest x-rays with deep learning, arXiv preprint arXiv:1711.05225
  (2017).

\bibitem{liu2017tx}
C.~Liu, Y.~Cao, M.~Alcantara, B.~Liu, M.~Brunette, J.~Peinado, W.~Curioso,
  Tx-cnn: Detecting tuberculosis in chest x-ray images using convolutional
  neural network, in: 2017 IEEE International Conference on Image Processing
  (ICIP), IEEE, 2017, pp. 2314--2318.

\bibitem{breiman1996stacked}
L.~Breiman, Stacked regressions, Machine learning 24~(1) (1996) 49--64.

\bibitem{kingma2014adam}
D.~P. Kingma, J.~Ba, Adam: A method for stochastic optimization, arXiv preprint
  arXiv:1412.6980 (2014).

\bibitem{wolpert1992stacked}
D.~H. Wolpert, Stacked generalization, Neural networks 5~(2) (1992) 241--259.

\bibitem{scikit-learn}
F.~Pedregosa, G.~Varoquaux, A.~Gramfort, V.~Michel, B.~Thirion, O.~Grisel,
  M.~Blondel, P.~Prettenhofer, R.~Weiss, V.~Dubourg, J.~Vanderplas, A.~Passos,
  D.~Cournapeau, M.~Brucher, M.~Perrot, E.~Duchesnay,
  \href{https://scikit-learn.org/stable/modules/generated/sklearn.multiclass.OneVsRestClassifier.html}{Scikit-learn:
  Machine learning in {P}ython}, Journal of Machine Learning Research 12 (2011)
  2825--2830.
\newline\urlprefix\url{https://scikit-learn.org/stable/modules/generated/sklearn.multiclass.OneVsRestClassifier.html}

\end{thebibliography}

\end{document}